\documentclass[aps,amssymb,prl,twocolumn,showpacs]{revtex4}
\usepackage{epsfig}
\usepackage{array}

\def\0{\varnothing}

\begin{document}
\title{Griffiths singularities in unbinding of strongly disordered polymers}
\author{Y. Kafri and D. Mukamel}
\affiliation{Department of Physics of Complex Systems, Weizmann
Institute of Science, Rehovot, Israel 76100.}
\date{November 21, 2002}

\begin{abstract}
Griffiths singularities occurring in the unbinding of strongly
disordered heteropolymers are studied. A model with two randomly
distributed binding energies $-1$ and $-v$, is introduced and
studied analytically by analyzing the Lee-Yang zeros of the
partition sum. It is demonstrated that in the limit $v \to \infty$
the model exhibits a Griffiths type singularity at a temperature
$T_G =O(1)$ corresponding to melting of long homogeneous domains
of the low binding energy. For finite $v \gg 1$ the model is
expected to exhibit an additional, unbinding, transition at a high
temperature $T_M=O(v)$.

\end{abstract}


\maketitle

The unbinding transition of two polymers has attracted
considerable attention in recent years. This phenomenon arises in
a variety of physical contexts in which two linearly extended
objects interact. Examples are the denaturation transition of
double stranded DNA \cite{WB} where the two strands of the
molecule unbind; wetting in two dimensions \cite{Fisher84} where
the linear interface separating two bulk phases becomes unbound
from the substrate; the depinning of flux lines from a columnar
defect in type II superconductors \cite{Nelson} and others. In
these problems an attractive interaction binds the two polymers,
or the other linear objects, to each other at low temperatures.
However, as the temperature is increased to a critical temperature
a transition takes place at which they become unbound.

In the case of homopolymers the unbinding transition is fairly
well characterized. Using either a directed polymer approach
\cite{Lipowsky} or the more general Poland-Scheraga (PS) type
models \cite{PS2} the nature of the transition and the critical
behavior have been studied. In the PS approach, introduced in the
context of DNA denaturation, a microscopic configuration of the
two polymers is viewed as an alternating sequence of bound
segments and denaturated loops. A statistical weight is assigned
to each bound segment and open loop. The weight of a bound segment
of length $l$ is taken as $w^l$ with $w=\exp(-\beta \epsilon)$,
$\beta=1/(k_B T)$ being the inverse temperature, and $\epsilon$ is
the binding energy of two monomers. On the other hand a loop of
length $l$ is assumed to carry no energy but rather a degeneracy
factor which takes the form $\Omega(l) \sim s^l/l^c$. Here $s$ is
a geometrical non-universal constant while $c$ is a universal
constant.

The nature of the unbinding transition was found to depend only on
the parameter $c$ \cite{PS2}. The transition is continuous for $1
< c \leq 2$ and is first order for $c>2$. No phase transition
takes place for $c \leq 1$ and the polymers are bound at all
temperatures. The value of $c$ taken in the PS approach is
universal and depends only on the dimension $d$ of the system and
on wether or not long range interactions, such as self-avoiding
interactions, are taken into account \cite{PS2,Fisher,Kafri}. In
the directed polymer approach in $d+1$ dimensions $c$ is related
to the number of random walks which return to the origin for the
first time. This is known to be given by~ \cite{Redner} $c=2-d/2$
for $d<2$, $c=d/2$ for $d>2$. In $d=2$ one has $\Omega(l)\sim
s^l/(l \ln^2 l)$.

In many cases one is interested in the unbinding transition of two
heteroploymers which are composed of a sequence of two or more
types of monomers. This is the case, for example, in DNA where
each strand is composed of a sequence of four types of
nucleotides. Heteropolymer unbinding is usually modelled by
introducing a quenched disorder in the contact energy between the
two chains. This disorder, besides modifying the nature of the
unbinding transition, may introduce Griffiths singularities in the
free energy as is the case in, for example, random
magnets~\cite{Grif}. In this Letter we study these singularities
in strongly disordered heteropolymer unbinding.

For weak disorder the Harris criterion~\cite{Harris} may be
applied in order to determine wether or not the critical behavior
of the homopolymer unbinding is modified by the disorder. One
finds that weak disorder is irrelevant as long as
$c<3/2$~\cite{For,For2,Derrida}. This corresponds to $1<d<3$ for
directed polymers. The marginal $c=3/2$ case, corresponding to
$d=1$ and $d=3$ in directed polymers, has been studied in
detail~\cite{For2,Derrida,Bhat,Lass,CH} and is found to be
marginally relevant~\cite{Derrida,Bhat,Lass}.

Recently the case of strong disorder has been considered by Tang
and Chat\'{e} \cite{Tang}. In this study the unbinding transition
is analyzed using a real-space renormalization group scheme and
numerical studies using a transfer matrix approach. It is
suggested that in the strong disorder regime the unbinding
transition becomes very smooth, of a Kosterlitz-Thouless (KT)
type, with a weak free energy singularity of the form $\exp(-a/
\sqrt{\vert T-T_M \vert})$ for $T<T_M$. The free energy is
differentiable to all orders at $T=T_M$. On the other hand, an
analysis of the transition by Azbel \cite{Azbel} suggests that the
transition could be of very high but finite order (over $100$).
The order of the transition depends on non-universal quantities
like the details of the distribution of the binding energies.

In this Letter we introduce a simple model for the unbinding
process of heteropolymers in the strong disorder regime. We
demonstrate that the free energy exhibits a Griffiths type
singularity \cite{Grif,Wor,Har,Sch} at a low temperature, $T_G$,
resulting from the melting of long homogeneous domains of sites
with low binding energy. The model then exhibits an unbinding
transition at a higher temperature, $T_M$.

The model considered in this study is a PS model with two types of
randomly distributed binding sites. The energy $\epsilon_i$ of
site $i$ is given by:
\begin{equation}
\epsilon_i = \left\{ \begin{array}{lcl} -1 &,& {\rm with \;probability}\;  p \nonumber\\
\nonumber
\\-v &,& {\rm with \; probability} \; 1-p\;.
\end{array}\right.
\label{eq:model}
\end{equation}
We first study the limit $v \to \infty$, where no unbinding takes
place. We demonstrate that a Griffiths singularity occurs at a
temperature $T_G=O(1)$. We then argue that the transition persists
at finite but large $v$, where it is followed by an unbinding
transition at a higher temperature $T_M=O(v)$. This model is
similar in spirit to a class of models used in the past to study,
for example, random field Ising models \cite{Mukamel}.

The simplifying feature of the limit $v \to \infty$ is that the
chain is decomposed into independent non-interacting finite
segments. The free energy of the model is thus obtained by
properly averaging the free energy of finite segments, each of
which is homogeneous. The free energy density of the chain may
thus be expressed as
\begin{equation}
{\cal F}=(1-p)^2 \sum_{N=0}^{\infty} p^N f_N(t) \;.
\label{eq:avefree}
\end{equation}
Here $t=(T_G-T)/T_G$, where $T_G$ is the melting temperature of a
homopolymer with $\epsilon=-1$. The free energy density $f_N(t)$
is that of a homogeneous chain of length $N+2$ where the first and
last sites of the two stands are taken to be bound. In the
thermodynamic limit the singular part of the free energy is given
by \cite{Fisher}
\begin{equation}
\lim_{N \to \infty} f_N(t) \;\sim \left\{ \begin{array}{lcl} 0 &,& \;t<0 \nonumber\\
\nonumber
\\t^\nu &,& \;t>0\;,
\end{array}\right.
\label{eq:fedensity}
\end{equation}
where $\nu$ is the critical exponent associated with the
divergence of the correlation length $\xi \sim t^{-\nu}$. The
exponent is given by \cite{Fisher}
\begin{equation}
\nu = \left\{ \begin{array}{lcl} 1/(c-1) &,& {\rm for}\;  1 < c \leq 2 \nonumber\\
\nonumber \\
1 &,& {\rm for} \; c>2 \;.
\end{array}\right.
\label{eq:nu}
\end{equation}
Clearly, at any finite $N$, $f_N(t)$ is an analytic function of
$t$ and the singularity develops only in the limit $N \to \infty$.
The sum (\ref{eq:avefree}) is a canonical case where Griffiths
type singularities may arise.

We proceed by first demonstrating that the average free energy
(\ref{eq:avefree}) is singular at $t=0$. This is done by showing
that $t=0$ is an accumulation point of the poles of the free
energy in the complex $t$ plane. The average free energy is then
proved to be differentiable to all orders at the transition.

\begin{figure}
\centerline{\epsfig{file=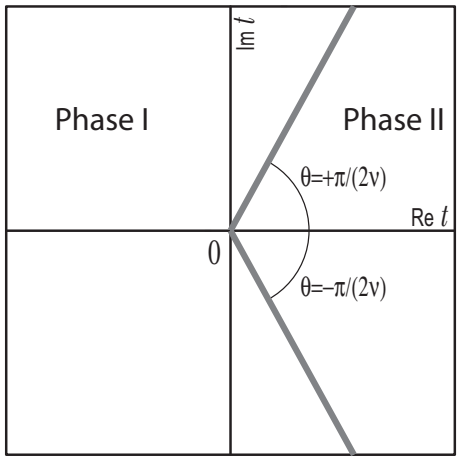,width=6truecm}} \caption{A
schematic illustration of the lines (in gray) in the complex $t$
plane on which the Lee-Yang zeros of the partition sum lie. The
density of zeros on these lines satisfies $\rho \sim r^{\nu-1}$. }
\label{fig:LYzeros}
\end{figure}

To show that the free energy (\ref{eq:avefree}) is singular at
$t=0$, we use the Lee-Yang theory \cite{LY} and analyze $f_N(t)$
in the complex $w=\exp(\beta)$ plane for large $N$. The partition
sum $Z_N(w)$ of a finite chain is a polynomial of order $N$ or
less in the complex $w=w_R+iw_I$ plane. It has no zeros on the
positive real $w$ axis. Let $\rho_N(w_R,w_I)$ be the density of
zeros in the complex $w$ plane. The free energy density $f_N(w)$
may be expressed as
\begin{equation}
f_N(w)=-\frac{1}{\beta N}\int \rho(x,y) \ln(w-z) dx dy
\label{eq:LYfe}
\end{equation}
where $z=x+iy$. To calculate $\rho(x,y)$ we note that it can be
expressed as \cite{LY,Der}
\begin{equation}
\rho(x,y)=2 \pi \nabla^2 \phi\;, \label{eq:LYden}
\end{equation}
with
\begin{equation}
\phi=\lim_{N \to \infty} \frac{\ln \vert Z_N \vert}{N} \;.
\end{equation}
One may thus view $-\phi$ as an electrostatic potential in a
two-dimensional space generated by a charge density $\rho(x,y)$.
We proceed by calculating $\phi$ near the transition where $w-w_G
\propto t$. It is evident from Eq. (\ref{eq:fedensity}) that the
system exhibits two phases: an unbound phase (I) near the real
negative $t$ axis where $Z_N=1$ and a bound phase (II) near the
real positive $t$ axis where $Z_N=\exp(Nt^\nu)$. In the complex
$t$ plane the potential in the two phases is
\begin{equation}
\phi(t) = \left\{ \begin{array}{lcl} 0  &,& {\rm phase \;\; I} \nonumber\\
\nonumber
\\ r^\nu \cos(\theta \nu) &,& {\rm phase \;\; II}\;,
\end{array}\right.
\label{eq:phases}
\end{equation}
where $t=r \exp(i \theta)$. The boundary between the two phases is
found by requiring that $\phi$ is continuous. This yields the
lines $\theta= \pm \pi/(2\nu)$. Hence phase II exists inside the
wedge $-\pi/(2 \nu) < \theta < +\pi/(2 \nu)$ while phase I exists
on the outside (see Fig. \ref{fig:LYzeros}). The discontinuity in
the field $\vec{E}=\nabla \phi$ in the direction perpendicular to
the phase boundary yields the charge density and thus the density
of zeros of the partition sum on that line. The angular component
of the electric field $E_\theta=1/r
\partial_\theta \phi$ in phase II on the transition lines $\theta= \pm
\pi/(2\nu)$ is given by
\begin{equation}
E_\theta=\mp \nu r^{\nu-1} \;.
\end{equation}
Thus, the density of the zeros of the partition sum along these
lines in the $N \to \infty$ limit is
\begin{equation}
\rho(r)=\frac{\nu}{2 \pi} r^{\nu-1} \;.
\end{equation}
It is evident that for $c>2$, where $\nu=1$, the density of zeros
is constant. Thus the typical distance $r_0$ to the first zero is
of order $r_0 \sim1/N$. For $1<c<2$, where $\nu=1/(c-1)$, the
density of zeros vanishes as $t \to 0$. The typical distance,
$r_0$, is determined by $\int_0^{r_0} r^{\nu-1} dr \sim 1/N$. This
yields $r_0 \sim 1/N^{c-1}$. Note that this distance goes to zero
in the thermodynamic limit as long as $c>1$. On the other hand for
$c<1$, $r_0$ does not vanish in the thermodynamic limit and thus
no singularity develops in the partition sum and no transition
takes place, as expected.

\begin{figure}
\centerline{\epsfig{file=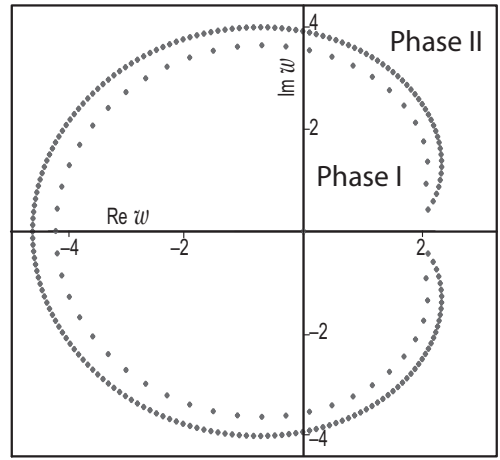,width=6.5truecm}}
\caption{The Lee-Yang zeros of the partition sum (\ref{eq:c3/2})
for a direct walk model corresponding to $c=3/2$ are shown for
$N=100$ (inner points) and $N=380$ (outer points). }
\label{fig:zeros}
\end{figure}

We have checked this picture by numerically evaluating the zeros
of the partition sum of a simple model of a directed polymer in
$1+1$ dimension corresponding to $c=3/2$. In this model we
consider a random walk which at each step moves either up ($x\to
x+1$) or down ($ x\to x-1$) with probability $1/2$. We consider
walks of length $N$ such that $x_0=x_N=0$ for which $x \geq 0$ at
all intermediate steps. With each visit to the origin, $x=0$, we
associate a weight $w$. The partition sum of this model is
\begin{equation}
Z_N=\sum_{R=1}^{N/2} w^R Q_{N,R}\;,\label{eq:c3/2}
\end{equation}
where $Q_{N,R}=\frac{R}{N-R} {N-R \choose N/2}$ is the number of
such walks of length $N$ which return to the origin $R$ times
\cite{Feller}. It is straightforward to show that in the $N \to
\infty$ limit, the model undergoes a transition at $w=2$. The
zeros of the partition sum (\ref{eq:c3/2}) have been calculated
numerically for $N=100$ and $N=380$ and are shown in Fig.
\ref{fig:zeros}. It is readily seen that the zeros are located on
a curve in the complex $w$ plane. As $N$ increases this curve
develops a wedge near $w=2$. In the large $N$ limit the curve
should approach the point $w=2$ at an angle $\theta=\pm \pi/4$.

We now turn to the average free energy density ${\cal F}$, Eq.
\ref{eq:avefree}. It is clear that $t=0$ is an accumulation point
of the poles of this free energy (or equivalently of the zeros of
the corresponding partition sum). Thus the series expansion of the
free energy at $t=0$ has a zero radius of convergence and the free
energy is singular. However, this singularity is expected to be
rather weak since the weight of the poles located at a distance
$O(1/N^{1/\nu})$ from $t=0$ (which correspond to $f_N(t)$ in
(\ref{eq:avefree})) have an exponentially small weight $p^N$. This
is a typical case of a Griffiths singularity.

To demonstrate that the average free energy ${\cal F}(t)$ is
differentiable to all order at $t=0$ we adopt the approach of
\cite{Sch} used for analyzing the Griffiths singularities in
random ferromagnets. We start by considering the $m$th derivative
of $f_N(t)$
\begin{eqnarray}
&& \hspace{-8mm}\frac{\partial^m f_N}{\partial \beta^m}=
\\&&\hspace{-8mm}-\frac{1}{\beta N}\sum_{\{k_i\}}
A(k_1,k_2,\ldots,k_l) \langle \vert E \vert^{k_1} \rangle \langle
\vert E \vert^{k_2} \rangle \ldots \langle \vert E \vert^{k_l}
\rangle \;, \nonumber
\end{eqnarray}
where the sum over $\{ k_i \}$ is taken with the constraint
$k_1+k_2+\ldots+k_l=m$ and $E$ is the energy of a chain of length
$N+2$ with the first and last sites bound. The brackets denote a
thermal average. The coefficients $A(k_1,k_2,\ldots,k_l)$ are
combinatorial factors independent of the system size $N$. Since
the energy at each site is either $-1$ (bound site) or $0$
(unbound site), one has on the real $t$ axis $\langle \vert E
\vert^{k} \rangle \leq N^k$. Thus
\begin{equation}
\left\vert \frac{\partial^m f_N}{\partial \beta^m} \right\vert
\leq \frac{N^{m-1}}{\beta} \left\vert \sum_{\{k_i\}}
A(k_1,k_2,\ldots,k_l) \right\vert \;.
\end{equation}
This inequality may be written as
\begin{equation}
\left\vert \frac{\partial^m f_N}{\partial \beta^m} \right\vert
\leq \frac{1}{\beta}g(m) N^{m-1} \label{eq:bound}
\end{equation}
where $g(m)$ is independent of $N$ and $\beta$. We now consider
the $m$th derivative of the average free energy ${\cal F}$,
\begin{equation}
\frac{\partial^m {\cal F}}{\partial \beta^m}  = (1-p)^2
\sum_{N=0}^\infty p^N \left( \frac{\partial^m f_N}{\partial
\beta^m} \right) \;.
\end{equation}
It is evident from (\ref{eq:bound}) that this series converges
absolutely which allows us to interchange the order of the
derivative and the sum. In fact, on the real $t$ axis the $m$th
derivative of ${\cal F}$ is bounded by
\begin{eqnarray}
\left\vert \frac{\partial^m {\cal F}}{\partial \beta^m}
\right\vert &\leq& (1-p)^2g(m) \sum_{N=1}^\infty p^N N^{m-1}
\nonumber \\
&\sim& (1-p)^2 g(m) \frac{\Gamma(m)}{\vert \ln p \,\vert^m} \;,
\end{eqnarray}
where $\Gamma(m)$ is the Gamma function. Therefore the derivative
exists on the real $t$ axis and particularly at $t=0$. This
analysis implies that the average free energy is differentiable to
all order in $\beta$ at $t=0$.

An insight of the singularity of the free energy may be obtained
by considering only the poles closest to $t=0$ in $f_N(t)$ for
calculating ${\cal F}$. For example for $c>2$ these poles are
located at $t=\pm ib/N$, where $b$ is a constant. Approximating
the sum on $N$ in (\ref{eq:avefree}) by an integral and after
rescaling $t$ one finds
\begin{equation}
{\cal F} \propto -\frac{1}{\beta} \int_0^\infty e^{-x}
\ln(x^2t^2+1) dx \;.
\end{equation}
This integral is singular at $t=0$ as expected.

So far we analyzed the limit $v \to \infty$, where no unbinding
takes place. Let us now consider finite $v$ with $v \gg 1$. In
this case the model is expected to exhibit an unbinding transition
at $T_M = O(v)$, in addition to the Griffiths singularity at
$T_G$. This may be seen by noting that at low temperatures, $T
\geq T_G$, the two polymers are expected to be bound even for
finite but large $v$. The reason is that the free energy
difference between the state which is bound at all strongly
binding sites and the completely unbound state is
\begin{equation}
\Delta F=-pvL+TpLu \ln s \;,
\end{equation}
where $u$ is a constant of $O(1)$, $\ln s$ is the entropy density
of a random walk and $L$ is the polymer length. Clearly the
second, entropic, term is negligibly small at $T \ll v$. Thus the
polymers are expected to be bound at low temperatures with an
unbinding transition taking place at $T_M =O(v)$.

The fact that $v$ is finite but large is not expected to modify
the nature of the Griffiths singularity found for $v \to \infty$
at $T=T_G$. At finite $v$ some of the strongly binding sites may
become unbound even at the low temperature around $T_G$. However
this effect is small and is not expected to modify the exponential
behavior of the domain size distribution taken in
(\ref{eq:avefree}). For $T>T_G$ where all homogeneous domains are
unbound our model is closely related to the Tang-Chat\'{e} model
in which the melting transition has been analyzed. Their analysis
of the melting transition should be applicable to the melting
transition in this model as well.

In summary, we have demonstrated that the model considered in this
study exhibits a Griffith singularity in the limit of infinitely
strong binding sites $v \to \infty$. It is argued that this
transition is present even for finite but large $v$, where it is
followed at a higher temperature by an unbinding transition.

\begin{acknowledgments}
We thank E. Domany, Y. Frishman and D. Kandel for helpful
discussions.
\end{acknowledgments}

\end{document}